\documentclass[structabstract]{aa}
\usepackage{graphicx}
\usepackage{natbib}
\newcommand{\chir}{\ensuremath{\chi_\nu^{\,2}}}                   % Reduced chi-squared symbol

\usepackage{amsmath,amssymb,graphicx}
\begin{document}

\title{Physical properties of the HAT-P-23 and WASP-48 planetary systems from multi-colour photometry}
%   \subtitle{}
\titlerunning{Physical properties of HAT-P-23\,b \& WASP-48\,b}

   \author{
          S. Ciceri\inst{1}
          \and
          L. Mancini \inst{1}
          \and
          J. Southworth\inst{2}
          \and
          I. Bruni\inst{3}
          \and
          N. Nikolov\inst{4}
          \and
          G. D'Ago\inst{5,6}
          \and \\
          T. Schr\"{o}der\inst{1}
          \and
          V. Bozza\inst{5,6}
          \and
          J. Tregloan-Reed\inst{7}
          \and
          Th. Henning\inst{1}
          }
{
       \institute{Max Planck Institute for Astronomy, K\"{o}nigstuhl 17, 69117 -- Heidelberg, Germany \\
             \email{ciceri@mpia.de}
         \and
    Astrophysics Group, Keele University, Staffordshire, ST5 5BG, UK
        \and
    INAF -- Osservatorio Astronomico di Bologna, Via Ranzani 1, 40127 -- Bologna, Italy
        \and
    Astrophysics Group, University of Exeter, Stocker Road, EX4 4QL, Exeter, UK
        \and
    Department of Physics, University of Salerno, Via Ponte Don Melillo, 84084 -- Fisciano (SA), Italy
        \and
    Istituto Nazionale di Fisica Nucleare, Sezione di Napoli, Italy
        \and
    NASA Ames Research Center, Moffett Field, CA 94035, USA
}

%   \date{Received ; Accepted}
 \abstract
 % 5 {} token are mandatory
{Accurate and repeated photometric follow-up observations of planetary-transit events are important to precisely characterize the physical properties of exoplanets. A good knowledge of the main characteristics of the exoplanets is fundamental to trace their origin and evolution. Multi-band photometric observations play an important role in this process.
}
{By using new photometric data, we computed precise estimates of the physical properties of two transiting planetary systems at equilibrium temperature of $\sim2000$\,K.}
{We present new broad-band, multi-colour, photometric observations obtained using three small class telescopes and the telescope-defocussing technique. In particular we obtained 11 and 10 light curves covering 8 and 7 transits of HAT-P-23 and WASP-48 respectively. For each of the two targets, one transit event was simultaneously observed through four optical filters.
One transit of WASP-48\,b was monitored with two telescopes from the same observatory. The physical parameters of the systems were obtained by fitting the transit light curves with {{\sc jktebop}} and from published spectroscopic measurements.}
{We have revised the physical parameters of the two planetary systems, finding a smaller radius for both HAT-P-23\,b and WASP-48\,b, $R_{\mathrm{b}}=1.224 \pm 0.037 \, R_{\mathrm{Jup}}$ and $R_{\mathrm{b}}=1.396 \pm 0.051 \, R_{\mathrm{Jup}}$, respectively, than those measured in the discovery papers ($R_{\mathrm{b}}=1.368 \pm 0.090 \, R_{\mathrm{Jup}}$ and $R_{\mathrm{b}}=1.67 \pm 0.10 \, R_{\mathrm{Jup}}$). The density of the two planets are higher than those previously published ($\rho_{\mathrm{b}}$ $\sim 1.1$ and $\sim 0.3 $ $\rho_{\mathrm{jup}}$ for HAT-P-23 and WASP-48 respectively) hence the two Hot Jupiters are no longer located in a parameter space region of highly inflated planets. An analysis of the variation of the planet's measured radius as a function of optical wavelength reveals flat transmission spectra within the experimental uncertainties. We also confirm the presence of the eclipsing contact binary NSVS-3071474 in the same field of view of WASP-48, for which we refine the value of the period to be $0.459$\,d.}
{}


\maketitle
\begin{figure}%
\centering
\includegraphics[width=\columnwidth]{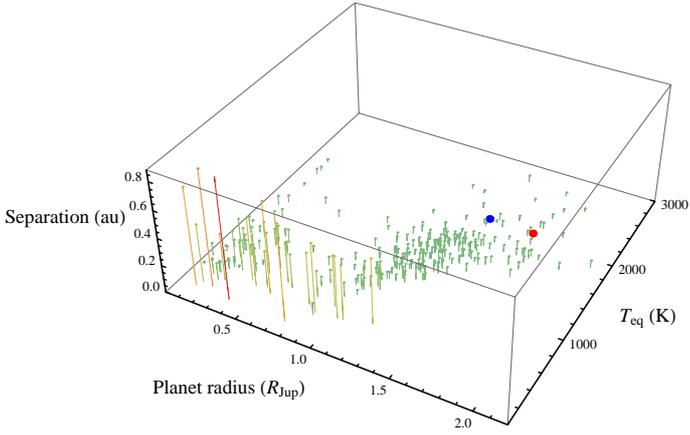}
\caption{3D plot of the known transiting exoplanets. The quantities on the axes are the planetary
radius and the temperature, and their semi-major axes. The positions of HAT-P-23\,b and WASP-48\,b
are highlighted by a blue and red point, respectively. Data taken from the TEPcat catalogue.}
\label{fig3d_lc}
\end{figure}


% Sect. 1
%%%%%%%%%%%%%%%%%%%%%%%%%%%%%%%%%%%%%%%%%%%%%%%%%%%%%%
\section{Introduction}
\label{sec_1}
%%%%%%%%%%%%%%%%%%%%%%%%%%%%%%%%%%%%%%%%%%%%%%%%%%%%%%

Among the almost 2000 extrasolar planets known to exist, those that transit their parent stars are of particular interest. In contrast to the exoplanets detected with other techniques, most of the physical and orbital parameters of transiting extrasolar planet (TEP) systems are measurable with precisions of a few percent using standard astronomical methods (e.g \citealp{seager2003, sozzetti2007, southworth2007, torres2008, southworth2008}). Obtaining estimations of the planets' masses and sizes can give some hint and direction in discriminating between gaseous and rocky structure and therefore infer their formation and evolution history. 
In particular, precise measurements of planetary sizes can provide strong constraints for those theoretical models that try to explain the inflation mechanisms for highly irradiated gaseous planets. 
Indeed, after the discovery of a group of inflated planets (e.g. WASP-17b, TrES-4b and HAT-P-32, \citealp{anderson2011, sozzetti2015, seeliger2014}), several theories invoking, tidal friction, enhanced atmospheric opacities, turbulent mixing, ohmic dissipation, windshocks or more exotic mechanisms, have been proposed to account for the slow cooling rate of these planets, resulting in a so unexpected large radius (see e.g. \citealp{baraffe2014, spiegel2013, ginzburg2015} and references therein).
A deduction of their chemical composition is also possible by looking for elemental and molecular signatures in transmission spectra observed during transit events \citep{seager2000,brown2001}.

In this context, we are carrying out a large program to study these TEP systems and robustly determine their physical properties via photometric monitoring of transit events in different passbands. We are utilising an array of 1-2 meter class telescopes, located in both of Earth's hemispheres, and single or multi-channel imaging instruments. In some cases, a two-site observational strategy was adopted for simultaneous follow-up of transit events (e.g.\ \citealp{ciceri2013,mancini2013a,mancini2014a}). We have so far refined the measured parameters of several TEP systems (e.g.\ \citealp{southworth2011wasp7,mancini2013a,mancini2014c}), studied starspot crossing events \citep{mancini2013c,mancini2014b}, and probed opacity-induced variations of measured planetary radius with wavelength (e.g.\ \citealp{southworth2012al,mancini2013b,mancini2013c,mancini2014b,nikolov2013}).

In this work, we focus our attention on the HAT-P-23 and WASP-48 systems, both hosting a star with effective temperature $T_{\mathrm{eff}}\sim6000$\,K and a close-in gas-giant transiting planet with a high equilibrium temperature of $T_{\mathrm{eq}}\sim 2000$\,K}. We have reported these characteristics in Fig.\,\ref{fig3d_lc}, together with those of the other known TEPs (data taken from TEPCat\footnote{TEPcat is the catalogue of the physical properties of transiting planetary systems \citep{southworth2011} and is available at http://www.astro.keele.ac.uk/jkt/tepcat/.}). The other main properties of the two TEP systems are summarized in the next two subsections. Observations of 15 new transit events of the two planets, performed at two different observatories, are presented in Sect.\,\ref{sec_2} together with the reduction of the corresponding photometric data. The analysis of the light curves, described in Sect.\,\ref{sec_3}, is followed by the revision of the physical parameters of the two planetary systems in Sect.\,\ref{sec_4}. We also present new light curves of the eclipsing binary NSVS 3071474, which is located close to WASP-48. Our conclusions are summarized in Sect.\,\ref{sec_5}.

% Figure 00

% Sect. 1.1
%%%%%%%%%%%%%%%%%%%%%%%%%%%%%%%%%%%%%%%%%%%%%%%%%%%%%%
\subsection{HAT-P-23}
\label{sec_1.1}
%%%%%%%%%%%%%%%%%%%%%%%%%%%%%%%%%%%%%%%%%%%%%%%%%%%%%%

The HAT-P-23 system, discovered by \citet{bakos2011}, is composed of a G0 dwarf star (mass $1.13\, M_{\sun}$, radius $1.20\, R_{\sun}$ and metallicity $+0.13$) and a hot Jupiter (mass $2.09\, M_{\mathrm{Jup}}$ and radius $1.37\, R_{\mathrm{Jup}}$), revolving around its parent star on a circular orbit with a period of $P=1.2$\,d. Using the SOPHIE spectrograph, \citet{moutou2011} measured the projected spin-orbit angle through observation of the Rossiter-McLaughlin (R-M) effect. Their finding of $\lambda=15^{\circ} \pm 22^{\circ}$ suggests an aligned and prograde planetary orbit. A reanalysis of the main parameters of the system was performed by \citet{ramon2013}. Recently, \citet{orourke2014} reported accurate photometry of planet occultations observed at $3.6 \, \mu$m and $4.5 \, \mu$m with the \emph{Spitzer} space telescope and at $H$ and $K_{\mathrm{S}}$ bands with the Hale Telescope. They found the emission spectrum to be consistent with a planetary atmosphere having a low efficiency of energy transport from its day-side to night-side, no thermal inversion and a lack of strongly-absorbing substances, similar to the case of WASP-19\,b \citep{mancini2013c}.

% Sect. 1.2
%%%%%%%%%%%%%%%%%%%%%%%%%%%%%%%%%%%%%%%%%%%%%%%%%%%%%%
\subsection{WASP-48}
\label{sec_1.2}
%%%%%%%%%%%%%%%%%%%%%%%%%%%%%%%%%%%%%%%%%%%%%%%%%%%%%%

WASP-48 is a TEP system composed of a slightly evolved F-type star (mass $1.19\, M_{\sun}$, radius $1.75\, R_{\sun}$ and metallicity $-0.12$) and an inflated hot Jupiter (mass $0.98\, M_{\mathrm{Jup}}$, radius $1.67\, R_{\mathrm{Jup}}$ and an orbital period of $P=2.1$\,d) \citep{enoch2011}. The lithium abundance and the absence of Ca\,II H and K emission suggests that the host star is old and evolving off the main sequence. However, by measuring the age from the rotation period, the star seems to be much younger \citep{enoch2011}. This discrepancy can be explained if tidal forces between the planet and the parent star have spun up the latter, making it difficult to obtain a reasonable value for the age based on gyrochronology (e.g.\ \citealp{pont2009}). The emission spectrum of WASP-48\,b was also investigated by \citet{orourke2014} through IR observations of the planet occultations  with the \emph{Spitzer} and the Hale telescopes. The nature of the atmosphere of WASP-48\,b was found to be very similar (moderate energy recirculation, no temperature inversion, absence of strong absorbers) to that of HAT-P-23\,b.

%  Sect. 2
%%%%%%%%%%%%%%%%%%%%%%%%%%%%%%%%%%%%%%%%%%%%%%%%%%%%%%
\section{Observations and data reduction}
\label{sec_2}
%%%%%%%%%%%%%%%%%%%%%%%%%%%%%%%%%%%%%%%%%%%%%%%%%%%%%%

% Table 1
\begin{table*}
\centering
\label{ObsLog}
\caption{Details of the observations presented in this work. $N_{\rm obs}$ is the number of observations, $T_{\rm exp}$ is the
exposure time, $T_{\rm obs}$ is the mean observational cadence, and `Moon illum.' is the fractional illumination of the Moon.}
% \tiny %
\setlength{\tabcolsep}{4pt}
\begin{tabular}{lccccccccccc}
\hline
Telescope & Date of   & Start time & End time  &$N_{\rm obs}$ & $T_{\rm exp}$ & $T_{\rm obs}$ & Filter & Airmass & Moon & Aperture   & Scatter \\
          & first obs &    (UT)    &   (UT)    &              & (s)           & (s)           &        &         &illum.& radii (px) & (mmag)  \\
\hline
\multicolumn{10}{l}{HAT-P-23:} \\
CA\,1.23-m & 2011 08 19 & 23:16 & 00:52 &  60 &     80     & 103  & Cousins $R$    & 1.78 $\to$ 1.34 & 67\% & 13,70,90 & 1.41 \\
CA\,1.23-m & 2012 07 21 & 00:34 & 04:30 & 139 &  50 to 120 & 101  & Cousins $R$    & 1.07 $\to$ 1.78 &  9\% & 14,70,90 & 0.69 \\
CA\,1.23-m & 2012 07 26 & 20:25 & 02:37 & 286 &  30 to 130 &  91  & Cousins $R$    & 1.70 $\to$ 1.07 & 58\% & 23,70,90 & 1.22 \\
CA\,1.23-m & 2013 06 16 & 22:50 & 03:49 & 120 &    135     & 151  & Cousins $R$    & 1.81 $\to$ 1.07 & 53\% & 23,60,90 & 0.95 \\
CA\,1.23-m & 2013 07 03 & 21:47 & 03:49 & 194 &  85 to 100 & 110  & Cousins $R$    & 1.78 $\to$ 1.07 & 15\% & 16,70,90 & 0.77 \\
CA\,1.23-m & 2013 07 26 & 20:25 & 02:37 &  87 &  30 to 130 &  61  & Cousins $R$    & 1.70 $\to$ 1.07 & 76\% & 23,70,90 & 1.59 \\
CA\,1.23-m & 2013 07 31 & 19:55 & 03:02 & 147 & 145 to 155 & 166  & Cousins $R$    & 1.80 $\to$ 1.07 & 28\% & 25,70,90 & 1.19 \\
CA\,2.2-m  & 2013 09 03 & 19:39 & 23:45 & 113 &  60 to 120 & 133  & Thuan-Gunn $u$ & 1.21 $\to$ 1.07 &  2\% & 10,25,40 & 3.55 \\
CA\,2.2-m  & 2013 09 03 & 19:39 & 23:45 & 117 &  60 to 120 & 133  & Thuan-Gunn $g$ & 1.21 $\to$ 1.07 &  2\% & 23,55,70 & 0.92 \\
CA\,2.2-m  & 2013 09 03 & 19:39 & 23:45 & 118 &  60 to 120 & 133  & Thuan-Gunn $r$ & 1.21 $\to$ 1.07 &  2\% & 21,45,65 & 0.71 \\
CA\,2.2-m  & 2013 09 03 & 19:39 & 23:45 & 116 &  60 to 120 & 133  & Thuan-Gunn $z$ & 1.21 $\to$ 1.07 &  2\% & 17,50,70 & 1.49 \\
\hline
\multicolumn{10}{l}{WASP-48:} \\
Cassini    & 2011 05 23 & 22:04 & 01:02 &  91 &  52 to  90 &  92  & Gunn $r$       & 1.30 $\to$ 1.04 & 57\% & 13,70,90 & 0.82 \\
Cassini    & 2011 05 25 & 21:33 & 02:45 & 222 &  50 to  90 &  81  & Gunn $r$       & 1.37 $\to$ 1.02 & 38\% & 13,70,90 & 0.81 \\
CA\,1.23-m & 2011 08 23 & 22:56 & 03:53 & 259 &  25 to  42 &  63  & Cousins $R$    & 1.77 $\to$ 1.27 & 29\% & 11,30,45 & 1.90 \\
CA\,1.23-m & 2012 09 12 & 19:08 & 01:25 & 136 & 120 to 160 & 151  & Cousins $I$    & 1.07 $\to$ 1.74 &  5\% & 20,70,90 & 1.01 \\
CA\,1.23-m & 2013 07 24 & 20:10 & 04:12 & 214 & 110 to 120 & 125  & Cousins $I$    & 1.05 $\to$ 1.59 & 92\% & 22,70,90 & 0.61 \\
CA\,2.2-m  & 2011 08 23 & 21:23 & 04:40 & 256 &  50 to  80 & 101  & Str\"omgren $u$ & 1.07 $\to$ 2.36 & 29\% & 10,16,30 & 2.88 \\
CA\,2.2-m  & 2011 08 23 & 21:23 & 04:40 & 246 &  50 to  80 & 101  & Thuan-Gunn $g$ & 1.07 $\to$ 2.36 & 29\% & 25,40,80 & 1.19 \\
CA\,2.2-m  & 2011 08 23 & 21:23 & 04:40 & 248 &  50 to  80 & 101  & Thuan-Gunn $r$ & 1.07 $\to$ 2.36 & 29\% & 23,33,60 & 1.09 \\
CA\,2.2-m  & 2011 08 23 & 21:23 & 04:40 & 134 &  50 to  80 & 101  & Cousins $I$    & 1.07 $\to$ 2.36 & 29\% & 18,28,50 & 1.25 \\
CA\,1.23-m & 2014 06 02 & 20:34 & 03:40 & 184 & 115 to 134 & 126  & Cousins $R$    & 2.12 $\to$ 1.05 & 25\% & 20,80,100& 0.89 \\
\hline
\end{tabular}
\end{table*}

In this section we present photometric observations of eight transits of HAT-P-23\,b and seven of WASP-48\,b. For both systems, one transit was observed with a multi-band imaging camera. One transit of WASP-48 was simultaneously monitored with two telescopes from the same observatory. The details of the observations are summarized in Table\,\ref{ObsLog}.

% Figure 01
\begin{figure}
\centering
\includegraphics[width=\columnwidth]{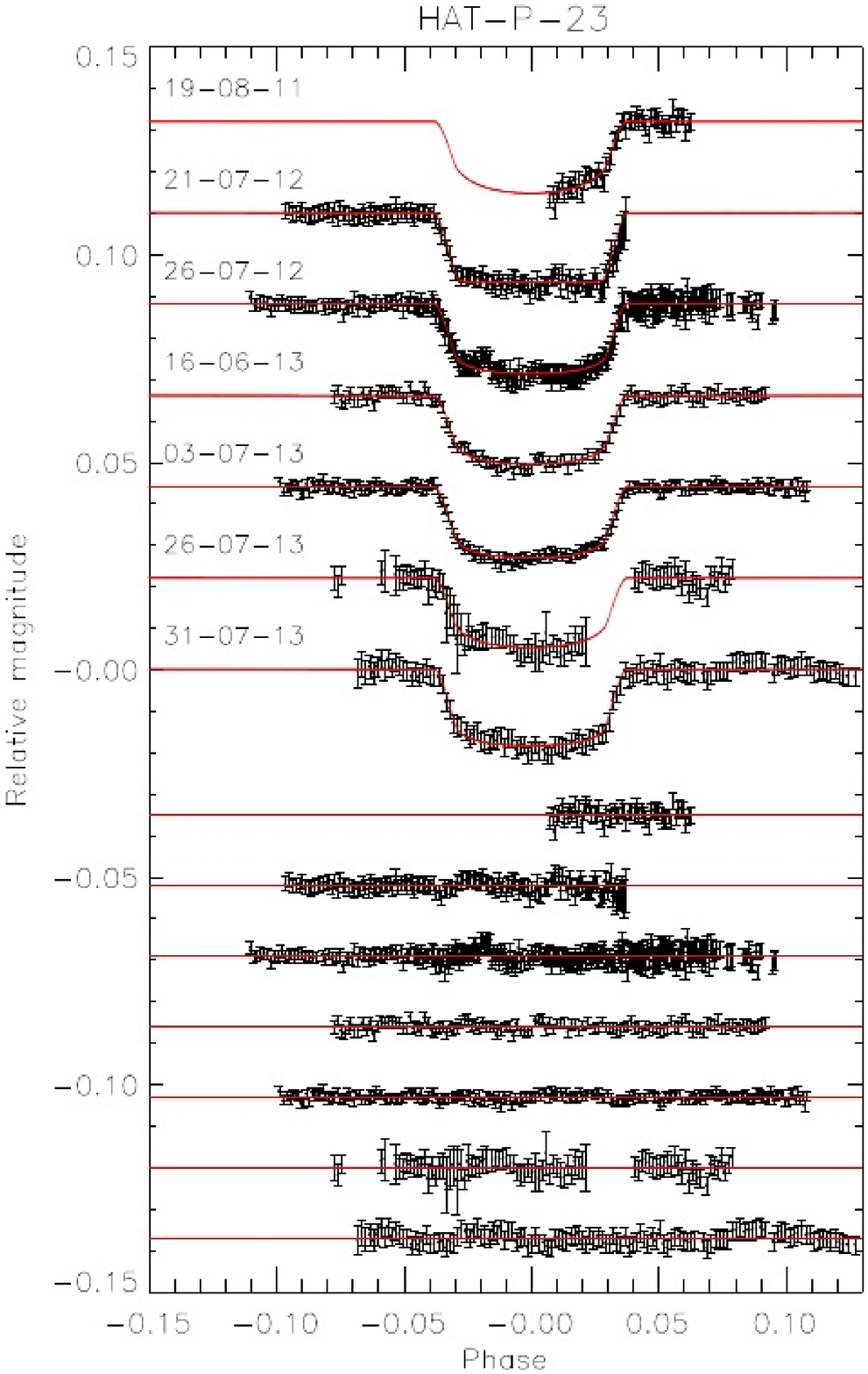}
\caption{Light curves of transits of HAT-P-23\,b, observed with the Calar Alto 1.23-m
telescope, compared with the best-fitting models given by {{\sc jktebop}}. The date of
each observation is reported next to the corresponding light curve. The residuals from
the fits are displayed at the base of the figure in the same order as the light curves.}
\label{hp-23_lc}
\end{figure}

% Figure 02
\begin{figure}
\centering
\includegraphics[width=\columnwidth]{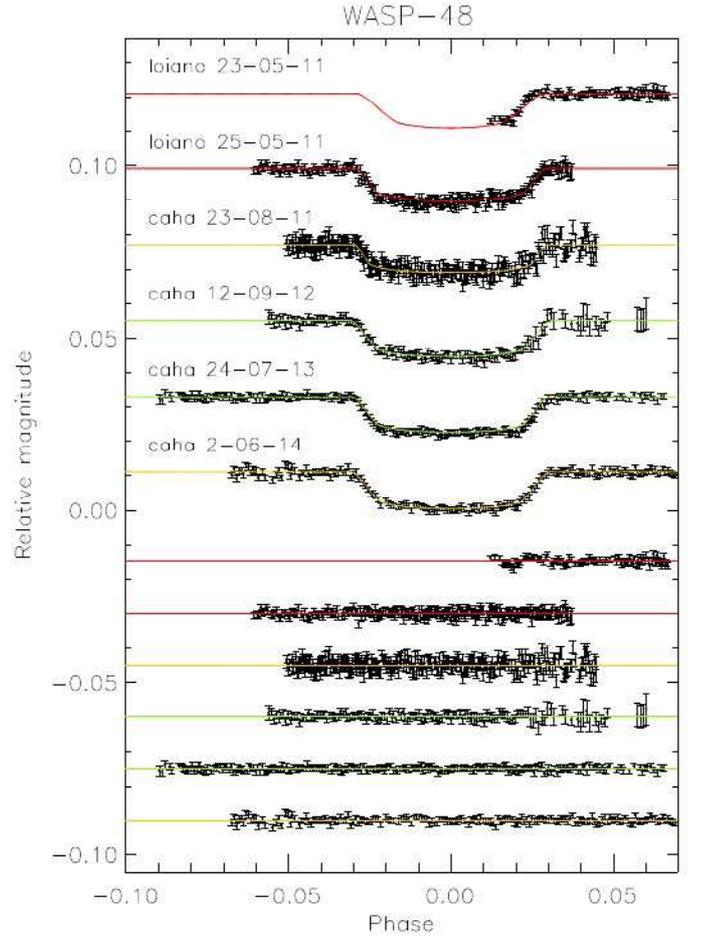}
\caption{Light curves of transits of WASP-48\,b, observed with the Cassini 1.52-m (``loiano'') and
Calar Alto 1.23-m (``caha'') telescopes, compared with the best-fitting curves given by {{\sc jktebop}}.
The date of each observation is reported next to the corresponding light curve. The residuals from
the fits are displayed at the base of the figure in the same order as the light curves.}
\label{w48_lc}
\end{figure}

% Figure 03   
\begin{figure}[!h]
\centering
\includegraphics[width=\columnwidth]{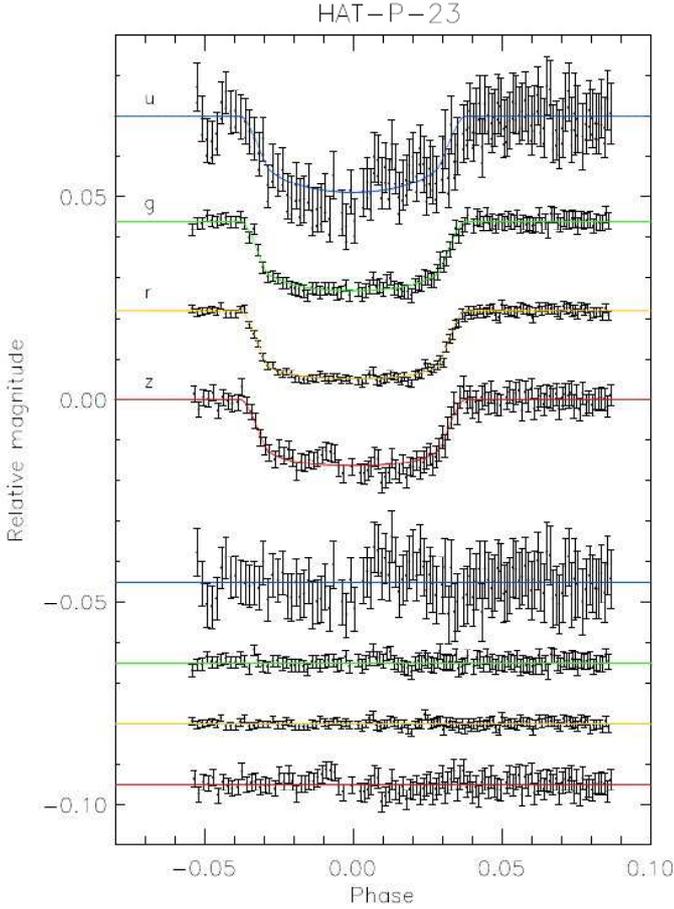}
\caption{A planetary transit event of HAT-P-23\,b as observed with the BUSCA instrument mounted on
the CAHA 2.2-m telescope. The light curves in the Thuan-Gunn $ugrz$ are shown from top to bottom and
are compared to the best {{\sc jktebop}} fits. The residuals are plotted at the base of the figure.}%
\label{busca_hp23}
\end{figure}

% Figure 04
\begin{figure}[!h]
\centering
\includegraphics[width=\columnwidth]{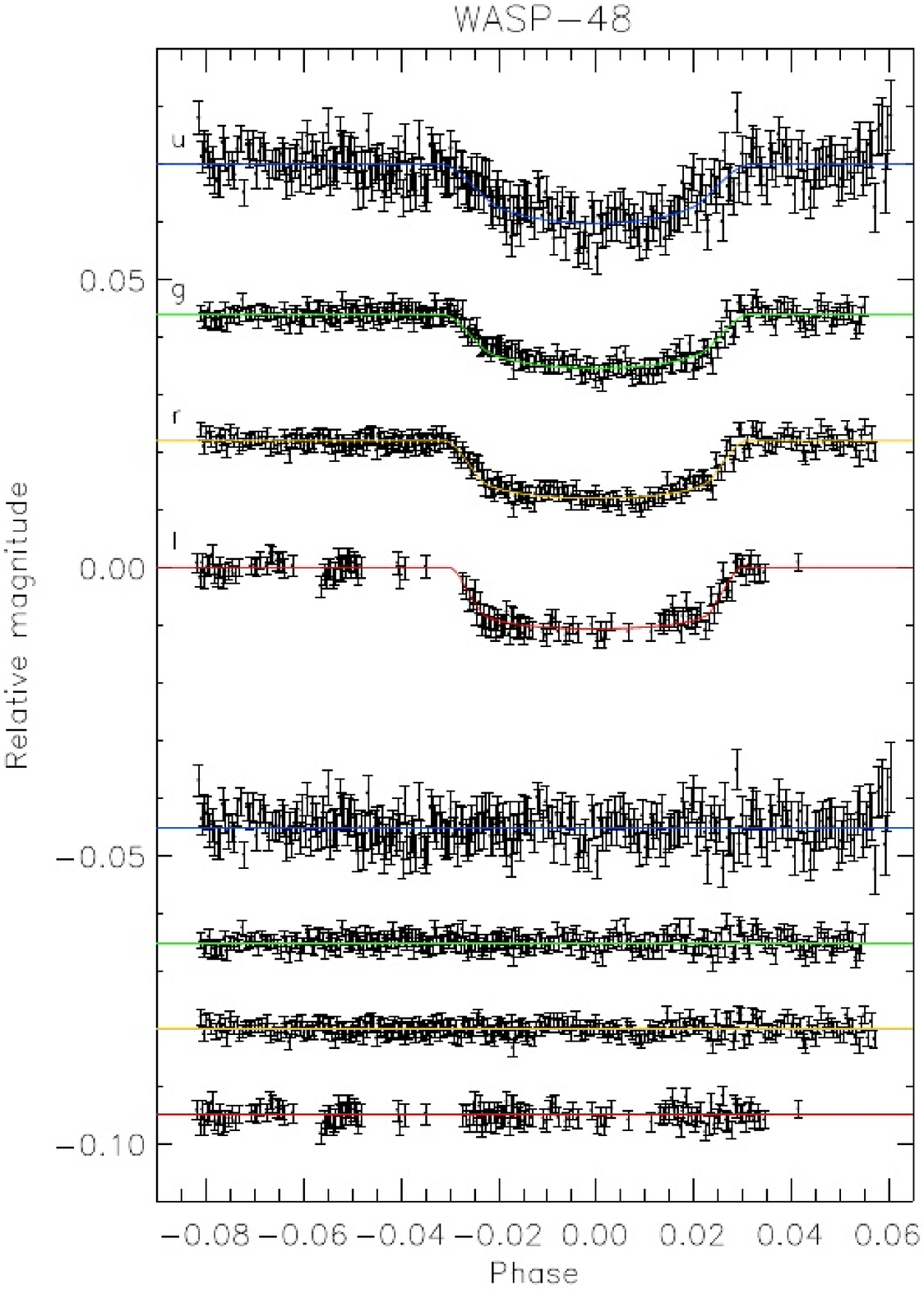}
\caption{As for Fig.\,\ref{busca_hp23} but for WASP-48\,b.
A Cousins $I$ filter was used in the reddest arm of BUSCA, while a Str\"omgren $u$ was used in the bluest one.}%
\label{busca_w48}
\end{figure}

%%%%%%%%%%%%%%%%%%%%%%%%%%%%%%%%%%%%%%%%%%%%%%%%%%%%%%
\subsection{Calar Alto 1.23-m Telescope}
\label{sec_2.1}
%%%%%%%%%%%%%%%%%%%%%%%%%%%%%%%%%%%%%%%%%%%%%%%%%%%%%%

Seven transits of HAT-P-23\,b and four of WASP-48\,b were remotely observed using the Zeiss 1.23-m telescope at the Calar Alto Observatory in Spain. The telescope has an equatorial mount and is equipped with a DLR-MKIII camera positioned at its Cassegrain focus. The CCD, which was used unbinned, has $4096 \times 4096$ pixels and a field of view (FOV) of $21.5^{\prime} \times 21.5^{\prime}$ leading to a resolution of $0.32^{\prime \prime}$ pixel$^{-1}$. The telescope was autoguided and defocussed for all science observations. This observing mode consists of using the telescope out of focus to spread the light of the stars in the FOV on many more pixels of the CCD than normal in-focus observations. In this way it is possible to use longer exposures, greatly increasing the signal to noise ratio (S/N) and reducing the uncertainties due to many sources of noise \citep{southworth2009al}. In our cases, the exposure times were fine-tuned at the beginning of each observation, together with the amount of defocussing, in order to properly optimize the S/N and have a maximum count per pixel for the target star between 25000 to 35000 ADUs. Once the defocussing amount was set, it was kept fixed for the entire monitoring of the transit (a typical PSF of the target covered a region with a diameter from 15 to 25 pixel). In some cases, it was necessary to modify the exposure time during the night to avoid the CCD saturation or to account for changes in counts caused by variation of the air mass of the target or weather conditions (e.g, dramatic variations of the external temperature or sudden appearance of cirrus and veils). The filter used to observe HAT-P-23\,b was Cousins $R$; for WASP-48\,b the first and last transits were observed through Cousins $R$ and the other two transits through Cousins $I$ (Table\,\ref{ObsLog}).

%%%%%%%%%%%%%%%%%%%%%%%%%%%%%%%%%%%%%%%%%%%%%%%%%%%%%%
\subsection{Cassini 1.52-m Telescope }
\label{sec_2.3}
%%%%%%%%%%%%%%%%%%%%%%%%%%%%%%%%%%%%%%%%%%%%%%%%%%%%%%

Two transit light curves of WASP-48 were obtained on 2011 May 23 and 25 at the Astronomical Observatory of Bologna in Loiano, Italy. Observations were carried out with BFOSC (Bologna Faint Object Spectrograph \& Camera) mounted at the Cassegrain focus of the 1.52-m Cassini telescope (see Table\,\ref{ObsLog}). The $1300 \times 1340$ pixels CCD has a FOV of $13^{\prime} \times 12.6^{\prime}$ resulting in a resolution of $0.58^{\prime \prime}$ pixel$^{-1}$. For both transits a Gunn-$r$ filter was used and the telescope was autoguided and defocussed. During the acquisition of the science images, the CCD was windowed to decrease the readout time, thus achieving a higher temporal cadence.

%%%%%%%%%%%%%%%%%%%%%%%%%%%%%%%%%%%%%%%%%%%%%%%%%%%%%%
\subsection{Calar Alto 2.2-m Telescope}
\label{sec_2.2}
%%%%%%%%%%%%%%%%%%%%%%%%%%%%%%%%%%%%%%%%%%%%%%%%%%%%%%

One transit of each object was observed with the Bonn University Simultaneous Camera (BUSCA, \citealp{reif1999}) mounted on the Calar Alto 2.2-m telescope. BUSCA can obtain photometry in four different passbands simultaneously, the incoming light being split by dichroics. Each of the four channels has a $4096 \times 4096$ pixel CCD, which were used with $2\times2$ binning to give a plate scale of $0.35^{\prime \prime}$ pixel$^{-1}$. The FOV of each channel depends on the filter in the light beam, being $5.8^{\prime}$ in diameter for the Thuan-Gunn and Str\"omgren filters and $12^{\prime} \times 12^{\prime}$ for the Cousins $I$ filter. The telescope was autoguided and defocussed during both observing sequences.

Unfortunately, the BUSCA controller requires the same exposure time to be used in all four channels. The exposure times were therefore chosen to avoid saturation in the $r$ band, for which the count rate was highest. This meant that the redder channels yielded better-quality data than the bluer channels, especially the $u$ band.

%%%%%%%%%%%%%%%%%%%%%%%%%%%%%%%%%%%%%%%%%%%%%%%%%%%%%%
\subsection{Data reduction}
\label{sec_2.4}
%%%%%%%%%%%%%%%%%%%%%%%%%%%%%%%%%%%%%%%%%%%%%%%%%%%%%%

Data reduction was performed using standard methods. Bias and flat-field images on the sky were collected before and during twilight, respectively, and median-combined to generate master bias and flat-field frames. These were used to calibrate the science images. Light curves were then obtained using aperture photometry algorithms from {\sc daophot} \citep{Stetson87pasp} as implemented in the {\sc idl}\footnote{The acronym {\sc idl} stands for Interactive Data Language and is a trademark of ITT Visual Information Solutions. For further details see: {\tt http://www.ittvis.com/ProductServices/IDL.aspx}.}-based {\sc defot} pipeline (see \citealt{southworth2014} and references therein), which uses subroutines from NASA's {\sc astrolib}\footnote{The {\sc astrolib} subroutine library is distributed by NASA. For further details see: {\tt http://idlastro.gsfc.nasa.gov/}.} library.

The sizes of the software apertures used for the aperture photometry are listed in Table\,\ref{ObsLog}, and were chosen to be those that gave the lowest out-of-transit (OOT) scatter. We noticed that changes in the aperture size of both the target region and the sky annulus do not affect the overall light-curve shape, but do cause a slight variation in the scatter of the data points. Once the aperture sizes were set, we extracted instrumental magnitudes for the target and possible comparison stars in the FOV. Pointing variations were corrected by cross-correlating each image against a reference frame. The light curves were then detrended by a second-order polynomial whilst optimising the weights of an ensemble of comparison stars. The choice of the comparison stars was carried out according to their brightness, by comparing the different OOT scatter and taking the combination that gave the lowest scatter.

%%%%%%%%%%%%%%%%%%%%%%%%%%%%%%%%%%%%%%%%%%%%%%%%%%%%%%
\section{Light curve analysis}
\label{sec_3}
%%%%%%%%%%%%%%%%%%%%%%%%%%%%%%%%%%%%%%%%%%%%%%%%%%%%%%

We separately analyzed each of the light curves observed using the {\sc jktebop}\footnote{The source code of \textsc{jktebop} is available at {\tt http:// www.astro.keele.ac.uk/jkt/codes/jktebop.html}} code \citep[see][and references therein]{southworth2008}. {\sc jktebop} fits an observed light curve with a synthetic one, constructed according to the values of a set of initial parameters. Some of the input parameters are left free to vary until the best fit is reached. The main photometric parameters that we can measure with {\sc jktebop} are the orbital inclination, $i$, the orbital period, $P$, the transit midpoint, $T_0$, and the sum and ratio of the fractional radii of the star and planet, $r_{\mathrm{A}}+r_{\mathrm{b}}$ and $k = r_{\mathrm{b}}/r_{\mathrm{A}}$. The fractional radii are defined as $r_{\mathrm{A}} = R_{\mathrm{A}}/a$ and $r_{\mathrm{b}} = R_{\mathrm{b}}/a$, where $a$ is the orbital semi-major axis, and $R_{\mathrm{A}}$ and $R_{\mathrm{b}}$ are the absolute radii of the star and the planet, respectively.

We assumed circular orbits for both the planetary systems \citep{enoch2011,orourke2014}. The values of the planet-star mass ratio were fixed to the those obtained using the estimated masses from the discovery papers. The limb darkening effect on the light curves was taken into account by modeling it using a quadratic law, and we checked that the difference in the results obtained using a linear or a logarithmic law is negligible.

The fits of the light curves were performed using theoretical values of the limb darkening coefficients \citep{claret2000,claret2004} and by fitting the linear coefficients whilst fixing the quadratic coefficients but perturbing them during the error analysis simulations. The atmospheric parameters of the stars HAT-P-23\,A and WASP-48\,A, assumed for deriving the initial values of the limb-darkening coefficients, are listed in Table\,\ref{ld-coeff}, as well as the weighted-mean values of the limb-darkening coefficient obtained from the fit of the light curves.

% Table 2 input 4 ld coeff
\begin{table}[!h]
\centering
\label{ld-coeff}
\caption{Stellar atmospheric parameters used to calculate the limb darkening coefficients, and weighted-mean values of the linear coefficients obtained from the fit of the light curves.}
\begin{tabular}{lcc}
\hline
Parameter & HAT-P-23\,A & WASP-48\,A  \\
\hline
$T_{\mathrm{eff}} (K)$             & 5905 & 6000  \\
$\log{g}$ (cm s$^{-2}$)            & 4.33 & 4.03  \\
$[\frac{\mathrm{Fe}}{\mathrm{H}}]$ & 0.1  & $-0.1$  \\
$V_{\mathrm{micro}}$ (km s$^{-1}$) & 2.0  & 2.0   \\
LD coeff $R$&$0.317\pm 0.031$ &$0.277\pm0.075$\\
LD coeff $I$&       -         &$0.101\pm0.067$\\
LD coeff $u$&$0.780\pm 0.511$ &$1.09\pm0.33$\\  
LD coeff $g$&$0.239\pm 0.090$ &$0.631\pm0.164$\\
LD coeff $r$&$0.221\pm 0.066$ &$0.229\pm0.065$\\
LD coeff $z$&$0.179\pm 0.139$ &       -       \\
\hline
\end{tabular}
\end{table}

Since the \textsc{aper} routine, which we used to perform aperture photometry on the calibrated science images, commonly underestimates the error bars, we enlarged them for each dataset by multiplying the errorbar for each photometric point by the square-root of the $\chir$ obtained through an initial fit of the corresponding light curve. We then further inflated the errors using the $\beta$ approach (e.g.\ \citealp{pont2006,gillon2006,winn2007}) to take into account the presence of systematic effects and correlated noise.

To assign the uncertainties to each parameter obtained from the fitting process, we generated $10\,000$ simulations with a Monte Carlo algorithm and also used a residual-permutation algorithm \citep{southworth2008}. Since none of the two techniques systematically gives lower uncertainties, we took the larger of the 1-$\sigma$ values obtained using the two algorithms. The best-fitting model and the residuals for each of the 21 light curves are shown in Figures \ref{hp-23_lc}, \ref{w48_lc}, \ref{busca_hp23}, and \ref{busca_w48}.

%%%%%%%%%%%%%%%%%%%%%%%%%%%%%%%%%%%%%%%%%%%%%%%%%%%%%%
\subsection{New orbital ephemeris}%
\label{sec_3.1}
%%%%%%%%%%%%%%%%%%%%%%%%%%%%%%%%%%%%%%%%%%%%%%%%%%%%%%

We refined the orbital periods of both planets by using our new photometric data and transit timings available in the literature or at the ETD\footnote{The Exoplanet Transit Database (ETD) website can be found at {\tt http://var2.astro.cz/ETD}} website.

We made a linear fit to the transit midpoints versus cycle numbers in order to improve the ephemeris. All the transits considered in the linear fit for the two planetary systems are listed in Tables \ref{res_hp23} and \ref{res_w48}. The new values for the ephemeris found from the fits are:
% hat-p-23
\begin{equation}
T_{0} = \mathrm{BJD(TDB)}\; 2\,454\,852.26599\,(20) + 1.21288287\,(17)\,E, \nonumber
\end{equation}
and
% wasp-48
\begin{equation}
T_{0} = \mathrm{BJD(TDB)}\; 2\,455\,364.55241\,(24) +2.14363544\,(58)\,E. \nonumber
\end{equation}
for HAT-P-23 and WASP-48, respectively. The numbers in brackets are the errors relative to the last digit, and $E$ represents the cycle number.

% Table 4
\begin{table}
\centering
\caption{Times of transit midpoint of HAT-P-23\,b and their residuals. References: (1) \cite{bakos2011}; (2) TRESCA; (3) CA 1.23-m, this work; (4) CA 2.2-m, this work.}%; (5) Salerno 35\,cm, this work.}
\label{res_hp23}
\tiny
\begin{tabular}{lrrl}
\hline
Time of minimum     & Epoch & Residual & Reference  \\
BJD(TDB) $-2400000$ &       &  (JD)    &            \\
\hline
$ 54852.265383 \pm 0.00018 $ &   0  &-0.00061 & (1) \\
$ 55434.453087 \pm 0.00103 $ &  480 & 0.00331 & (2) \\
$ 55451.430268 \pm 0.00112 $ &  494 & 0.00013 & (2) \\
$ 55736.460815 \pm 0.00086 $ &  729 & 0.00321 & (2) \\
$ 55749.802945 \pm 0.00072 $ &  740 & 0.00362 & (2) \\
$ 55753.439625 \pm 0.00112 $ &  743 & 0.00166 & (2) \\
$ 55776.484166 \pm 0.00085 $ &  762 & 0.00142 & (2) \\
$ 55783.762296 \pm 0.00051 $ &  768 & 0.00226 & (2) \\
$ 55812.871847 \pm 0.00114 $ &  792 & 0.00262 & (2) \\
$ 55838.341217 \pm 0.00063 $ &  813 & 0.00145 & (2) \\
$ 56129.433378 \pm 0.00060 $ & 1053 & 0.00172 & (2) \\
$ 56129.433908 \pm 0.00068 $ & 1053 & 0.00225 & (2) \\
$ 56129.432399 \pm 0.00041 $ & 1053 & 0.00057 & (3) \\
$ 56135.496648 \pm 0.00157 $ & 1058 & 0.00263 & (2) \\
$ 56135.495288 \pm 0.00032 $ & 1058 & 0.00158 & (3) \\
$ 56159.756359 \pm 0.00085 $ & 1078 & 0.00202 & (2) \\
$ 56186.438739 \pm 0.00090 $ & 1100 &-0.00075 & (2) \\
$ 56460.547674 \pm 0.00033 $ & 1326 & 0.00124 & (3) \\
$ 56477.528543 \pm 0.00025 $ & 1340 &-0.00137 & (3) \\
$ 56478.743946 \pm 0.00091 $ & 1341 & 0.00074 & (2) \\
$ 56500.574007 \pm 0.00281 $ & 1359 &-0.00079 & (3) \\
$ 56505.425061 \pm 0.00052 $ & 1363 &-0.00101 & (3) \\
$ 56539.383814 \pm 0.00168 $ & 1391 &-0.00050 & (4) $u$ \\
$ 56539.385859 \pm 0.00036 $ & 1391 & 0.00019 & (4) $g$ \\
$ 56539.385926 \pm 0.00021 $ & 1391 &-0.00029 & (4) $r$ \\
$ 56539.385145 \pm 0.00051 $ & 1391 &-0.00226 & (4) $z$ \\
$ 56562.430097 \pm 0.00122 $ & 1410 &-0.00021 & (2) \\
$ 56562.432087 \pm 0.00098 $ & 1410 &-0.00015 & (2) \\
$ 56579.409838 \pm 0.00114 $ & 1424 &-0.00093 & (2) \\
\hline
\end{tabular}
\end{table}

% Table 5
\begin{table}
\caption{Times of transit mid-point of WASP-48\,b and their residuals. References: (1) \cite{enoch2011}; (2) TRESCA; (3) Loiano 1.52-m, this work; (4) CA 1.23-m, this work; (5) CA 2.2-m, this work.}
\label{res_w48}
\centering
\tiny
\begin{tabular}{lrrl}
\hline
Time of minimum    & Epoch & Residual & Reference  \\
BJD(TDB)$-2400000$ &       &   (JD)   &            \\
\hline
$ 55364.55202 \pm 0.00027 $ &   0 &-0.00039 & (1) \\
$ 55696.80915 \pm 0.00131 $ & 155 &-0.00675 & (2) \\
$ 55696.81647 \pm 0.00155 $ & 155 & 0.00056 & (2) \\
$ 55750.41680 \pm 0.00151 $ & 180 & 0.01000 & (2) \\
$ 55782.56695 \pm 0.00195 $ & 195 & 0.00562 & (2) \\
$ 55797.56545 \pm 0.00160 $ & 202 &-0.00131 & (2) \\
$ 55825.43480 \pm 0.00152 $ & 215 & 0.00076 & (2) \\
$ 56033.37084 \pm 0.00162 $ & 312 & 0.00416 & (2) \\
$ 56065.51850 \pm 0.00187 $ & 327 &-0.00270 & (2) \\
$ 56168.42213 \pm 0.00067 $ & 375 & 0.00642 & (2) \\
$ 56391.34809 \pm 0.00218 $ & 479 &-0.00570 & (2) \\
$ 56393.49772 \pm 0.00087 $ & 480 & 0.00029 & (2) \\
$ 56453.51930 \pm 0.00181 $ & 508 & 0.00007 & (2) \\
$ 56468.53007 \pm 0.00110 $ & 515 & 0.00540 & (2) \\
$ 56487.81919 \pm 0.00060 $ & 524 & 0.00180 & (2) \\
$ 56511.40314 \pm 0.00252 $ & 535 & 0.00576 & (2) \\
$ 56541.40894 \pm 0.00103 $ & 549 & 0.00067 & (2) \\
$ 56541.41054 \pm 0.00139 $ & 549 & 0.00227 & (2) \\
$ 55705.37900 \pm 0.01013 $ & 159 &-0.01145 & (3) \\
$ 55707.53160 \pm 0.00041 $ & 160 &-0.00248 & (3) \\
$ 55797.57056 \pm 0.00063 $ & 202 & 0.00378 & (4) \\
$ 55797.56621 \pm 0.00162 $ & 202 &-0.00056 & (5) $u$ \\
$ 55797.56614 \pm 0.00063 $ & 202 &-0.00063 & (5) $g$ \\
$ 55797.56473 \pm 0.00053 $ & 202 &-0.00204 & (5) $r$ \\
$ 55797.56569 \pm 0.00050 $ & 202 &-0.00108 & (5) $I$ \\
$ 56183.42469 \pm 0.00062 $ & 382 & 0.00353 & (4) \\
$ 56498.53535 \pm 0.00025 $ & 529 &-0.00021 & (4) \\
$ 56811.50538 \pm 0.00025 $ & 675 &-0.00094 & (4) \\
\hline
\end{tabular}
\end{table}

If a TEP system, known to be composed of a parent star and a planet, hosts other planets, the gravitational interaction between the planetary components results in a periodic delay and advance in the times of transit of the known planet. We checked for a possible third component in the HAT-P-23 and WASP-48 systems by looking to see whether there was a periodic variation in the transit times of the known planets.

\begin{figure*}
\centering
\includegraphics[width=\textwidth]{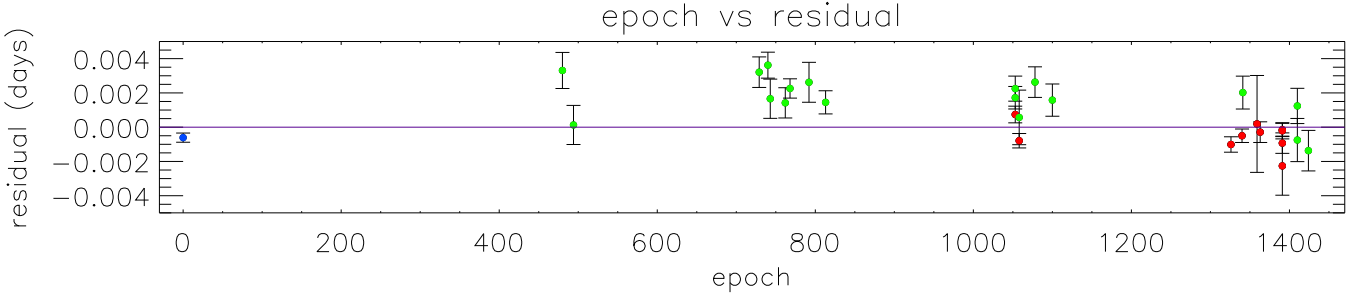}
\caption{Plot of the residuals of the timing of mid-transit of HAT-P-23 versus a linear ephemeris. The points indicate literature results (blue), data obtained from the TRESCA catalogue (red), and our data (green).}
\label{oc_hp23}
\end{figure*}

\begin{figure*}
\centering
\includegraphics[width=\textwidth]{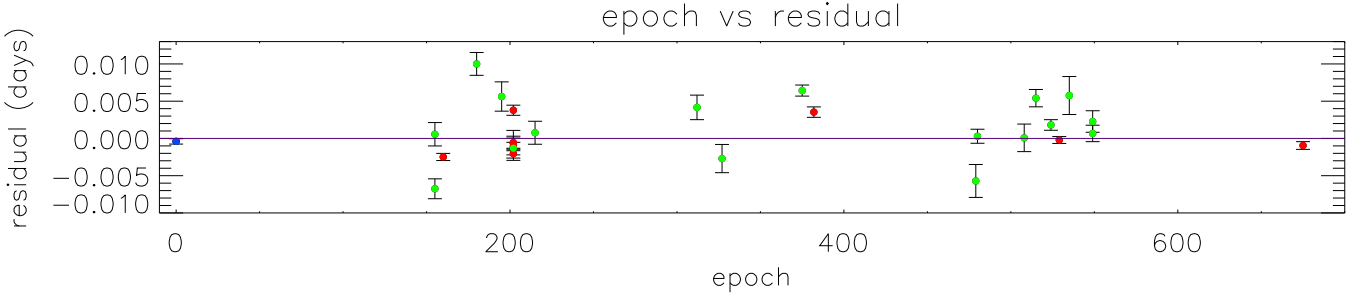}
\caption{Plot of the residuals of the timing of mid-transit of WASP-48 versus a linear ephemeris. The points indicate the value from the discovery paper (blue), data obtained from the TRESCA catalogue (red), and our data (green).}
\label{oc_w48}
\end{figure*}

The residuals from the linear fits are plotted in Figs.\ \ref{oc_hp23} and \ref{oc_w48} as a function of cycle number and do not show any clear systematic deviation from the predicted transit times. However, the quality of the fits, $\chi_{\nu}^2=3.26$ for HAT-P-23 and $\chi_{\nu}^2=10.05$ for WASP-48, indicates that a linear ephemeris is not a good match to the observations in both the cases. Based on our experience with a similar situation in previous studies (e.g.\ \citealp{southworth2012a,southworth2012al,mancini2013a,mancini2013c}), we conservatively do not interpret the large $\chi_{\nu}^2$ values as sign of transit timing variations, but as an underestimation of the uncertainties in the various $T_{0}$ measurements.

% Figure 05

% Figure 06

%%%%%%%%%%%%%%%%%%%%%%%%%%%%%%%%%%%%%%%%%%%%%%%%%%%%%%
\subsection{Final photometric parameters}
\label{sec_3.2}
%%%%%%%%%%%%%%%%%%%%%%%%%%%%%%%%%%%%%%%%%%%%%%%%%%%%%%

The final photometric parameters for the HAT-P-23 system were obtained taking into account only the complete transit light curves (i.e.\ discarding the two partial ones). This choice was made after noticing that the $\chir$, relative to the ratio of the radii of the planet and the star, decreases significantly, from $4.37$ to $0.99$, when rejecting them. Concerning WASP-48, we discarded the incomplete light curve and those with very high scatter (i.e.\ the light curve observed on 23-08-2011 with the CA 1.23-m and with BUSCA in the $u$ band). Even discarding three light curves, the $\chir$ for the $k$ parameter remained rather high, $8.48$. This discrepancy is discussed further in Section\,\ref{sec_4.1}.

The final values were obtained as a weighted mean of all the transits taken into account. The relative errors, obtained from the weighted mean, are rescaled by multiplying them for the relative $\chir$.

The parameters of the {\sc jktebop} fits to each of our light curves are given in Tables \ref{pho-res_hp23} and \ref{pho-res_w48} for HAT-P-23 and WASP-48, respectively. These tables also show the final photometric parameters in bold font, and the results from the discovery papers for comparison.

% Table *** HAT-P-23
\begin{table*}
\caption{Photometric properties of the HAT-P-23 system derived by fitting the light curves with
{\sc jktebop}. The final parameters are given in bold and are compared with those of the discovery paper. (*) Light curve not taken into account for the final result.}
\label{pho-res_hp23}
\centering
\begin{tabular}{lccccc}
\hline
Source & $r_{\mathrm{A}}+r_{\mathrm{b}}$& $k$ & $i$ & $r_{\mathrm{A}}$ &$r_{\mathrm{b}}$\\
\hline
CA\,1.23-m \# 1  *& $0.26447 \pm 0.02107$ & $0.11350 \pm 0.00267$ & $89.72 \pm 0.90$ & $0.23751 \pm 0.01893$ & $0.02696 \pm 0.00224$ \\
CA\,1.23-m \# 2  *& $0.28307 \pm 0.01135$ & $0.12220 \pm 0.00103$ & $82.01 \pm 1.19$ & $0.25225 \pm 0.01002$ & $0.03083 \pm 0.00138$ \\
CA\,1.23-m \# 3 & $0.23870 \pm 0.00861$ & $0.11343 \pm 0.00129$ & $88.48 \pm 2.23$ & $0.21438 \pm 0.00765$ & $0.02432 \pm 0.00102$ \\
CA\,1.23-m \# 4 & $0.24324 \pm 0.00856$ & $0.11196 \pm 0.00180$ & $87.05 \pm 2.44$ & $0.21875 \pm 0.00745$ & $0.02450 \pm 0.00116$ \\
CA\,1.23-m \# 5 & $0.25716 \pm 0.00625$ & $0.11732 \pm 0.00113$ & $84.48 \pm 0.92$ & $0.23016 \pm 0.00539$ & $0.02700 \pm 0.00086$ \\
CA\,1.23-m \# 6 & $0.25561 \pm 0.01734$ & $0.11175 \pm 0.00498$ & $86.56 \pm 3.42$ & $0.22992 \pm 0.01533$ & $0.02569 \pm 0.00216$ \\
CA\,1.23-m \# 7 & $0.25067 \pm 0.01582$ & $0.12028 \pm 0.00201$ & $85.94 \pm 3.22$ & $0.22376 \pm 0.01382$ & $0.02691 \pm 0.00205$ \\
CA\,2.2-m $u$   & $0.28919 \pm 0.03285$ & $0.12114 \pm 0.01224$ & $81.87 \pm 4.85$ & $0.25794 \pm 0.02940$ & $0.03125 \pm 0.00462$ \\
CA\,2.2-m $g$   & $0.25315 \pm 0.02248$ & $0.11856 \pm 0.00192$ & $84.18 \pm 3.71$ & $0.22632 \pm 0.01972$ & $0.02683 \pm 0.00278$ \\
CA\,2.2-m $r$   & $0.24223 \pm 0.00920$ & $0.11668 \pm 0.00121$ & $86.30 \pm 2.50$ & $0.21692 \pm 0.00819$ & $0.02531 \pm 0.00124$ \\
CA\,2.2-m $z$   & $0.23377 \pm 0.00981$ & $0.11505 \pm 0.00207$ & $87.17 \pm 2.64$ & $0.20965 \pm 0.00854$ & $0.02412 \pm 0.00131$ \\
\hline
{\bf Final results}  & $\mathbf{0.24539 \pm 0.00499}$ & $\mathbf{0.11616 \pm 0.00081}$ & $\mathbf{85.74 \pm 0.95}$ & $\mathbf{0.21998 \pm 0.00436}$ & $\mathbf{0.02541 \pm 0.00065}$ \\
\hline
\citet{bakos2011} & $0.268 \pm 0.014$ & $0.1169 \pm 0.0012$ & $85.1 \pm 1.5$ & $0.240 \pm 0.014$ & $0.028 \pm 0.014$ \\
\hline
\end{tabular}
\end{table*}

% Table *** WASP-48
\begin{table*}
\caption{Photometric properties of the WASP-48 system derived by fitting the light curves with {\sc jktebop}.
The final parameters are given in bold and are compared with those of the discovery paper. (*) Light curve not taken into account for the final result.}
\label{pho-res_w48}
\centering
\begin{tabular}{lccccc}
\hline
Source & $r_{\mathrm{A}}+r_{\mathrm{b}}$& $k$ & $i$ & $r_{\mathrm{A}}$ &$r_{\mathrm{b}}$ \\
\hline
Cassini \# 1     *& $0.23478 \pm 0.02433$ & $0.08349 \pm 0.00233$ & $84.33 \pm 2.50$ & $0.21668 \pm 0.02273$ & $0.01809 \pm 0.00184$ \\
Cassini \# 2      & $0.22734 \pm 0.01098$ & $0.09028 \pm 0.00081$ & $82.34 \pm 0.96$ & $0.20852 \pm 0.00995$ & $0.01883 \pm 0.00102$ \\
CA\,1.23-m \# 1  *& $0.18406 \pm 0.00280$ & $0.07810 \pm 0.00131$ & $89.90 \pm 0.65$ & $0.17072 \pm 0.00270$ & $0.01333 \pm 0.00022$ \\
CA\,1.23-m \# 2   & $0.23268 \pm 0.01670$ & $0.09426 \pm 0.00201$ & $82.49 \pm 1.70$ & $0.21264 \pm 0.01496$ & $0.02004 \pm 0.00179$ \\
CA\,1.23-m \# 3   & $0.23979 \pm 0.00790$ & $0.09648 \pm 0.00046$ & $81.56 \pm 0.67$ & $0.21870 \pm 0.00715$ & $0.02110 \pm 0.00074$ \\
CA\,2.2-m $u$    *& $0.28972 \pm 0.04443$ & $0.09503 \pm 0.01380$ & $78.88 \pm 3.66$ & $0.26458 \pm 0.03772$ & $0.02515 \pm 0.00678$ \\
CA\,2.2-m $g$     & $0.20267 \pm 0.01931$ & $0.08247 \pm 0.00407$ & $85.24 \pm 3.25$ & $0.18723 \pm 0.01750$ & $0.01544 \pm 0.00193$ \\
CA\,2.2-m $r$     & $0.22335 \pm 0.02172$ & $0.08900 \pm 0.00367$ & $83.23 \pm 2.25$ & $0.20509 \pm 0.01958$ & $0.01825 \pm 0.00230$ \\
CA\,2.2-m $I$     & $0.21368 \pm 0.01603$ & $0.09392 \pm 0.00145$ & $83.98 \pm 1.63$ & $0.19533 \pm 0.01449$ & $0.01835 \pm 0.00156$ \\
CA\,1.23-m \# 4   & $0.23689 \pm 0.01140$ & $0.09613 \pm 0.00126$ & $81.79 \pm 1.04$ & $0.21612 \pm 0.01020$ & $0.02078 \pm 0.00123$ \\
\hline
{\bf Final results} & $\mathbf{0.23276 \pm 0.00578}$ & $\mathbf{0.09584 \pm 0.00077}$ & $\mathbf{81.99 \pm 0.54}$ & $\mathbf{0.21272 \pm 0.00520}$ & $\mathbf{0.02010 \pm 0.00059}$ \\
\hline
\citet{enoch2011} & $0.259 \pm 0.013$ & $0.098 \pm 0.014$ & $80.09^{+0.88}_{-0.79}$ & $0.2364 \pm 0.0125$ & $0.0227 \pm 0.0272$ \\
\hline
\end{tabular}
\end{table*}

%%%%%%%%%%%%%%%%%%%%%%%%%%%%%%%%%%%%%%%%%%%%%%%%%%%%%%
\section{Physical properties of HAT-P-23 and WASP-48}
\label{sec_4}
%%%%%%%%%%%%%%%%%%%%%%%%%%%%%%%%%%%%%%%%%%%%%%%%%%%%%%

%Table 8
\begin{table}
\centering
\caption{\label{tab:spec} Spectroscopic parameters of the stars HAT-P-23\,A and WASP-48\,A.}
\label{init_par}
\begin{tabular}{lcccc}
\hline
Source      & HAT-P-23\,A & Ref. & WASP-48\,A & Ref. \\
\hline
$T_{\mathrm{eff}}$ (K)             & $5885 \pm 72$     & 1 & $6000 \pm 150$          & 4 \\
$[\frac{\mathrm{Fe}}{\mathrm{H}}]$ & $0.13 \pm 0.08$   & 1 & $-0.12 \pm 0.12$        & 4 \\
$K_{\rm A}$ (m\,s$^{-1}$)          & $368.5 \pm 17.6$  & 2 & $136.0^{+11.0}_{-11.1}$ & 4 \\
$e$                                & $0.0$ fixed       & 3 & $0.0$ fixed             & 4 \\
\hline
\end{tabular}
\tablefoot{(1) \citet{torres2012}; (2) \citet{bakos2011}; (3) \citet{orourke2014}; (4) \citet{enoch2011}. }
\end{table}

Following the methodology used by \citet{southworth2010}, the main physical parameters of the TEP systems HAT-P-23 and WASP-48 were found from the photometric parameters deduced from the parameters available in the literature (see Table\,\ref{init_par} for values and references), and by interpolating within the tabulated predictions of five sets of theoretical stellar models \citep{Claret04aa,Demarque+04apjs,Pietrinferni+04apj,Vandenberg++06apjs,Dotter+08apjs}. In brief, an initial estimate of the stellar mass was specified and the observed quantities were compared to the ones predicted by stellar models for this mass. The mass was then iteratively adjusted to find the best agreement between the observed and expected values.

Since the radius versus mass relation varies according to the age of the star, the interpolation was performed for different ages of the system, starting from 0.1\,Gyr until the end of the main-sequence lifetime of the star, in steps of 10\,Myr. The output set of physical parameters is the one that gives the best agreement between the predicted and the measured quantities. Separate sets of results were calculated using each of the five sets of theoretical model tabulations.

The final values found are shown in Table\,\ref{fin-res_hp23} for HAT-P-23 and Table\,\ref{fin-res_w48} for WASP-48. Most quantities have two errorbars, and in these cases the first is a statistical error obtained by propagating the uncertainties on the input measurements, and the second is a systematic error which is the scatter of the results from each of the five different sets of theoretical models (see \citealp{southworth2009}).

% Table 9
\begin{table}
\centering
\caption{Physical properties for the HAT-P-23 system. The results obtained in this work are reported and compared with those of the discovery paper.}
\label{fin-res_hp23}
\setlength{\tabcolsep}{2pt}
\begin{tabular}{lcc}
\hline
& This work (final) & \!\!\!\!Bakos et al.\ (2011) \\
\hline
$M_{\mathrm{A}}$ ($M_{\sun}$)               & $ 1.104 \pm 0.043 \pm 0.018$        & $ 1.13 \pm 0.04 $     \\ 
$R_{\mathrm{A}}$ ($R_{\sun}$)               & $ 1.089 \pm 0.027 \pm 0.006$        & $ 1.20 \pm 0.07 $     \\ 
$\log g_{\mathrm{A}}$ (cgs)                 & $ 4.407 \pm 0.018 \pm 0.002$        & $ 4.33 \pm 0.06 $ \\  
$\rho_{\mathrm{A}}$ ($\rho_{\sun}$)         & $ 0.855 \pm 0.051 $                 & $ - $                 \\ 
$M_{\mathrm{b}}$ ($M_{\mathrm{jup}}$)       & $ 2.07 \pm 0.12 \pm 0.02$           & $ 2.090 \pm0.111 $    \\ 
$R_{\mathrm{b}}$ ($R_{\mathrm{jup}}$)       & $ 1.224 \pm 0.036 \pm 0.007$        & $ 1.368 \pm 0.090 $   \\ 
$g_{\mathrm{b}}$ ($\mathrm{ms^{-2}}$)       & $ 34.3 \pm  2.4 $                   & $ 27.5 \pm 3.2 $      \\ 
$\rho_{\mathrm{b}}$ ($\rho_{\mathrm{jup}}$) & $ 1.057 \pm 0.097 \pm 0.006$        & $ 0.81 \pm 0.15 $     \\ 
$T_{\mathrm{eq}}$ ($\mathrm{K}$)            & $ 1951 \pm   30$                    & $ 2056 \pm 66 $       \\ 
$\Theta$                                    & $ 0.0706 \pm 0.0040 \pm 0.0004$     & $ 0.062 \pm 0.004 $   \\ 
$a$ (AU)                                    & $ 0.02302 \pm 0.00030 \pm 0.00012$  & $ 0.0232 \pm 0.0002 $ \\ 
Age (Gyr)                                   & $ 2.1_{-4.5 \,-2.0 }^{+3.4 \,+1.1 }$& $ 4.0 \pm 1.0 $       \\ 
\hline
\end{tabular}
\end{table}

% Table 10
\begin{table}
\caption{Final results for the physical parameters of WASP-48 obtained in this work compared to those of the discovery paper.}
\label{fin-res_w48}
\centering
\setlength{\tabcolsep}{2pt}
\begin{tabular}{lcc}
\hline
& This work (final) & \!\!\!\!Enoch et al.\ (2011) \\
\hline
$M_{\mathrm{A}}$ ($M_{\sun}$)               & $ 1.062  \pm0.074  \pm 0.014  $   & $ 1.19 \pm 0.05 $       \\
$R_{\mathrm{A}}$ ($R_{\sun}$)               & $ 1.519  \pm0.051  \pm 0.007  $   & $ 1.75 \pm 0.09 $       \\
$\log g_{\mathrm{A}}$ (cgs)                 & $ 4.101  \pm0.023  \pm 0.002  $   & $ 4.03 \pm 0.04 $       \\
$\rho_{\mathrm{A}}$ ($\rho_{\sun}$)         & $ 0.303  \pm0.022  $              & $ 0.22 \pm 0.03 $       \\
$M_{\mathrm{b}}$ ($M_{\mathrm{jup}}$)       & $ 0.907  \pm0.085  \pm 0.008  $   & $ 0.98 \pm 0.09 $       \\
$R_{\mathrm{b}}$ ($R_{\mathrm{jup}}$)       & $ 1.396  \pm0.051  \pm 0.006  $   & $ 1.67 \pm 0.10 $       \\
$g_{\mathrm{b}}$ ($\mathrm{ms^{-2}}$)       & $ 11.5   \pm1.1    $              & $ 8.1 \pm 1.1 $         \\
$\rho_{\mathrm{b}}$ ($\rho_{\mathrm{jup}}$) & $ 0.312  \pm0.037  \pm 0.001  $   & $ 0.21 \pm 0.04 $       \\
$T_{\mathrm{eq}}$ ($\mathrm{K}$)            & $ 1956   \pm54     $              & $ 2030 \pm 70 $         \\
$\Theta$                                    & $ 0.0406 \pm0.0036 \pm 0.0002 $   & $ - $                   \\
$a$ (AU)                                    & $ 0.03320\pm0.00077\pm 0.00015$   & $ 0.03444 \pm 0.00043 $ \\
Age (Gyr)                                   & $ 6.6_{-4.9 \,-2.8}^{+1.0 \,+0.6}$& $ 7.9_{-1.6}^{+2.0} $   \\
\hline
\end{tabular}
\end{table}

%%%%%%%%%%%%%%%%%%%%%%%%%%%%%%%%%%%%%%%%%%%%%%%%%%%%%%
\subsection{Radius vs wavelength variation}
\label{sec_4.1}
%%%%%%%%%%%%%%%%%%%%%%%%%%%%%%%%%%%%%%%%%%%%%%%%%%%%%%

Photometric observations of planetary transit events through different filters allow us to measure the apparent radius of transiting planets in each passband, obtaining an insight of the composition of their atmosphere (e.g \citealp{southworth2012al,copperwheat2013,nikolov2013,mancini2013c,narita2013,chen2014}).
We used our multi-band data to investigate possible variations of the radius of HAT-P-23\,b and WASP-48\,b in different optical passbands.

\begin{figure}
\centering
\includegraphics[width=\columnwidth]{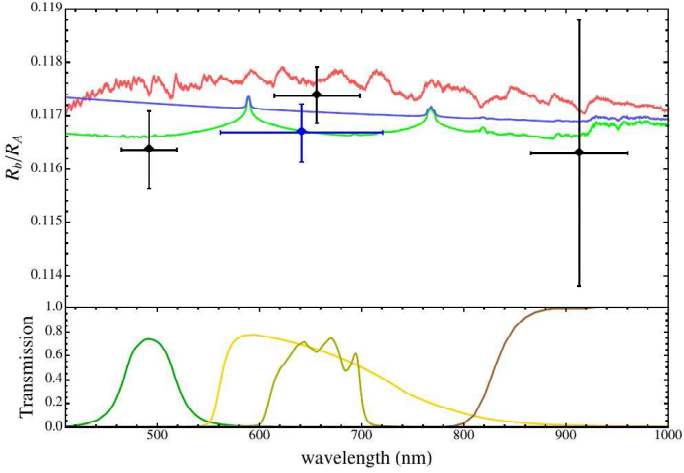}
\caption{Variation of the planetary radius of HAT-P-23\,b, in terms of planet/star radius ratio, with wavelength.
The black points are from the transit observed with BUSCA, while the blue point is the weighted-mean results coming
from the seven transits observed in Cousins $R$ with the CA\,1.23-m telescope. The vertical bars represent the errors
in the measurements and the horizontal bars show the FWHM transmission of the passbands used. The observational points
are compared with three synthetic spectra for a Jupiter planet with a surface gravity of $g_{\mathrm{b}}=25$\,m\,s$^{-2}$
and $T_{\mathrm{eq}}=2000$\,K. The synthetic spectra in green and blue do not include TiO and VO opacity, while the
spectrum in red does, based on equilibrium chemistry. With respect to the model identified with the green line, the blue
one has H$_{2}$/He Rayleigh scattering increased by a factor of 100. An offset is applied to all three models to provide
the best fit to our radius measurements. Transmission curves of the filters used are shown in the bottom panel.}
\label{hp23_wl-r}
\end{figure}

\begin{figure}
\centering
\includegraphics[width=\columnwidth]{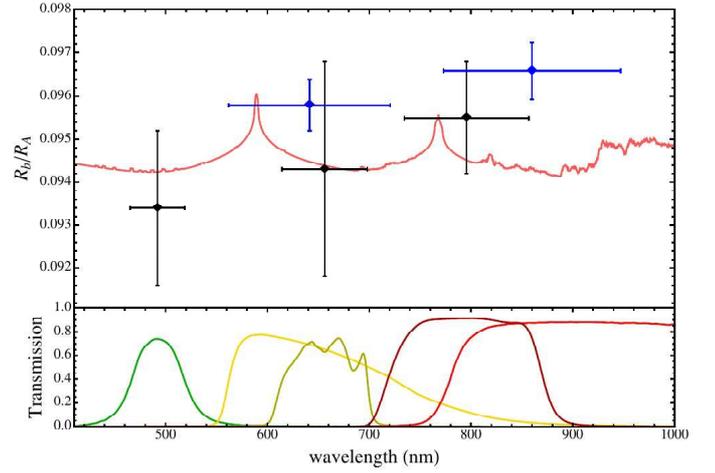}
\caption{Variation of the planetary radius of WASP-48\,b, in terms of planet/star radius ratio, with wavelength.
The black points are from the transit observed with BUSCA, while the blue points are the weighted-mean results
coming from the two transits observed in Cousins $R$ and the two in $I$ with the CA\,1.23-m telescope. The vertical
bars represent the errors in the measurements and the horizontal bars show the FWHM transmission of the passbands
used. The observational points are compared with a synthetic spectrum for a Jupiter planet with a surface gravity
of $g_{\mathrm{b}}=10$\,m\,s$^{-2}$ and $T_{\mathrm{eq}}=2000$\,K, which does not include TiO and VO opacity. An
offset is applied to all three models to provide the best fit to our radius measurements. Transmission curves of
the filters used are shown in the bottom panel.} \label{w48_wl-r}
\end{figure}

We phased all the light curves collected with the same telescope and filter combination and, following the strategy used by \citep{southworth2012al}, we re-fitted them and the BUSCA ones with all parameters fixed to the final values reported in Table\,\ref{fin-res_hp23} and \ref{fin-res_w48}, with the exception of $k$. This approach allows to remove sources of uncertainty common to all data sets, maximizing the accuracy of estimations of the planet/star radius ratio as a function of wavelength. Due to the very large uncertainty, the values of $k$ measured in the $u$ band were ignored. The results in the other bands are shown in Fig.\,\ref{hp23_wl-r} for HAT-P-23\,b and Fig.\,\ref{w48_wl-r} for WASP-48\,b. The vertical
bars represent the relative errors in the measurements and the horizontal bars show the full width at half-maximum (FWHM) transmission of the passbands used. Transmission curves of the adopted filters are shown in the bottom panel.

For illustration, the results obtained for HAT-P-23\,b are compared with three spectra calculated from one-dimensional model atmospheres by \citet{fortney2010} for a Jupiter-mass planet with a surface gravity of $g_{\mathrm{b}}=25$\,m\,s$^{-2}$, a base radius of $1.25 \, R_{\mathrm{Jup}}$ at 10\,bar, and $T_{\mathrm{eq}}=2000$\,K. The first model (red line) is run in an isothermal case taking into account chemical equilibrium and the presence of strong absorbers, like TiO and VO. The second model (green line) is obtained omitting the presence of the strong absorbers. The last model (blue line) is obtained by artificially removing the strong absorbers, as in the previous case, but also increasing the H$_{2}$/He Rayleigh scattering by a factor of 100. The values of $k$ for WASP-48\,b were compared with another model from \citet{fortney2010} similar to previous ones, but for a Jupiter-mass planet with a surface gravity of $g_{\mathrm{b}}=10$\,m\,s$^{-2}$ and without strong absorbers.

The precision of the final results does not allow us to discriminate among the models. We can only note that, since the are no large variations of the planets' radius at shorter wavelength within the experimental uncertainties, the atmospheres of the two planets should be quite transparent and not affected by large Rayleigh scattering. They are well-suited for further investigation at the optical wavelengths with more precise instruments.

% Figure 07

% Figure 08

%%%%%%%%%%%%%%%%%%%%%%%%%%%%%%%%%%%%%%%%%%%%%%%%%%%%%%
\subsection{Eclipsing Binary}
\label{appendix}
%%%%%%%%%%%%%%%%%%%%%%%%%%%%%%%%%%%%%%%%%%%%%%%%%%%%%%

The eclipsing binary candidate NSVS 3071474 is located at sky position RA(J2000) $=$ 19 24 03.821 and Dec(J2000) $=$ 55 27 33.33. It is sufficiently close to WASP-48 that it is present in the FOV of some of the transits that we monitored. We have extracted light curves of this object and confirm the nature of the system to be that of a contact eclipsing binary.

\begin{figure}
\caption{Light curves of NSVS 3071474 (1SWASP J192403.81+552734.5) phased (black dots) and binned
(red dots) observed by the WASP survey between 2008 and 2010.}
\label{eb-wasp}
\centering
\includegraphics[width=\columnwidth]{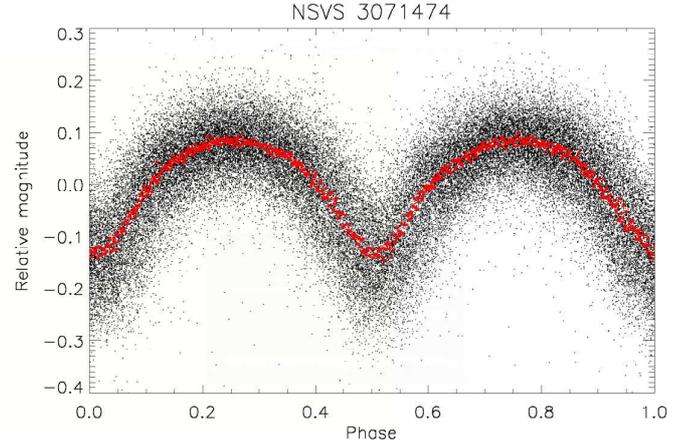}
\end{figure}

\begin{figure}
\caption{Light curves of NSVS 3071474, the eclipsing binary close to WASP-48.
The curves in the two top panels were observed with the 1.52-m Cassini telescope
while the two bottom ones were observed with the 1.23-m Calar Alto telescope.}
\label{eb}
\centering
\includegraphics[width=\columnwidth]{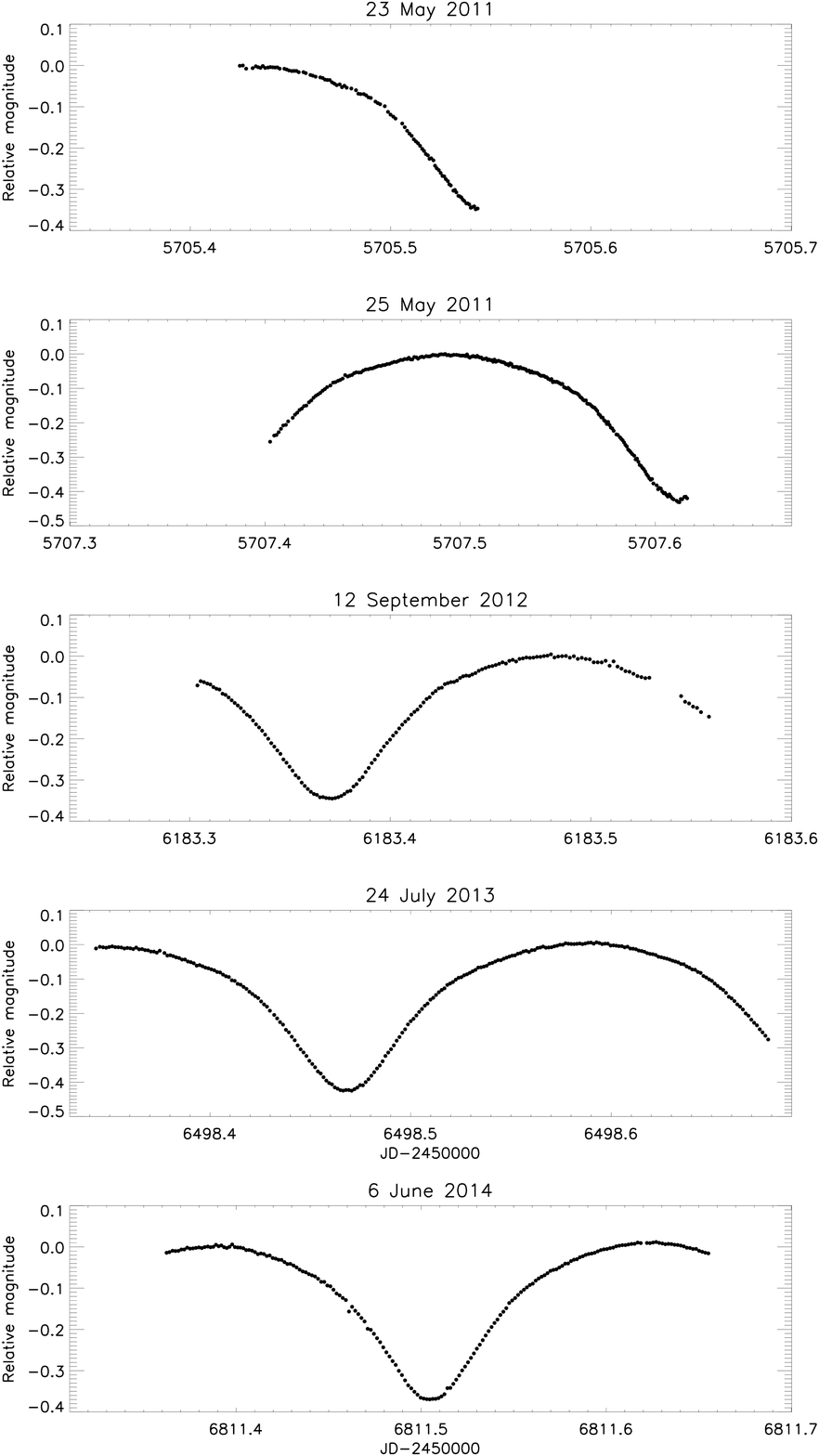}
\end{figure}

We also obtained the light curve observed by the SuperWASP survey \citep{pollacco2006}, where it is called 1SWASP J192403.81+552734.5. The phased and binned WASP data, and the light curves obtained from our frames are shown in Figs.\ \ref{eb-wasp} and \ref{eb}, respectively.

By fitting both the WASP data, and our lightcurves with JKTEBOP we obtained a measurement of the period of the eclipsing binary of $0.458998509 \pm 0.000000011$ days.

% Figure 09

% Figure 10

%%%%%%%%%%%%%%%%%%%%%%%%%%%%%%%%%%%%%%%%%%%%%%%%%%%%%%
\section{Discussion and Conclusions}
\label{sec_5}
%%%%%%%%%%%%%%%%%%%%%%%%%%%%%%%%%%%%%%%%%%%%%%%%%%%%%%

% Figure 11
\begin{figure}
\centering
\caption{Mass versus radius diagram of the transiting planets. The blue and red dots represent
the values for HAT-P-23 and WASP-48 found in literature and obtained in this work respectively.}
\label{r_vs_m}
\includegraphics[width=\columnwidth]{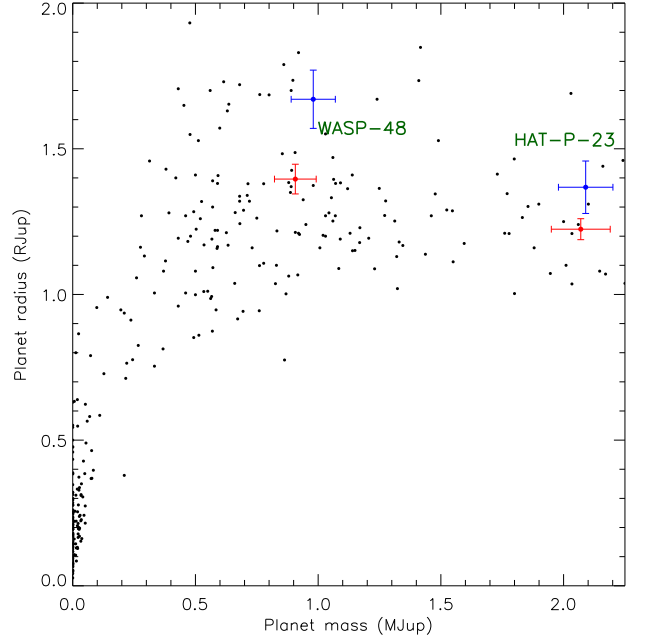}
\end{figure}

In this paper we presented refined parameters for the two TEP
systems HAT-P-23 and WASP-48, obtained from new photometric data
of transit events. We also presented simultaneous observations in
different optical bands. Our principal results are as follows.

\begin{itemize}

\item [$\bullet$] We confirmed the mass value of HAT-P-23\,b, but obtained a radius for it smaller by roughly 1.5$\sigma$. This TEP now occupies a more populated region of the mass-radius diagram and is no longer found to be one of the highly inflated transiting planets (see Fig.\,\ref{r_vs_m}). The mean density is therefore higher than what was found in the discovery paper.

\item [$\bullet$] We obtained improved estimates of the mass and radius of both the star and the planet of the WASP-48 planetary system. The values for the stellar and planetary radius are smaller than those found in the discovery paper by roughly 2.2$\sigma$ and 2.4$\sigma$, respectively. The masses are consistent with previous results.

\item [$\bullet$] A study of the planets' radius variation as a function of optical
wavelength, based on the data presented in this work, do not
indicate any large variation for either planet, suggesting that
their atmospheres are not affected by a large Rayleigh scattering.
Further investigations of the transmission spectra of these two
planets are needed to validate this statement.

\item [$\bullet$] Finally, we also presented new light curves for the eclipsing
binary NSVS\,3071474, refining the measurement of the orbital
period to be 0.458998509 (11) days.

\end{itemize}

\begin{acknowledgements}
Based on observations obtained with the 1.52-m Cassini telescope at the OAB Observatory in Loiano (Italy), and with the 1.23-m and 2.2m telescopes at the Centro Astron\'{o}mico Hispano Alem\'{a}n (CAHA) at Calar Alto (Spain), jointly operated by the Max-Planck Institut f\"{u}r Astronomie and the Instituto de Astrof\'{i}sica de Andaluc\'{i}a (CSIC). The reduced light curves presented in this work will be made available at the CDS (http://cdsweb.u-strasbg.fr/). Special thanks go to D.\ Pollacco and A.\ Cameron for allowing us to use both the public and the unpublished WASP data of the eclipsing binary NSVS\,3071474. S.C.\ thanks M.\ Line, J.\ Fortney and his group for useful discussions and for the kind hospitality at UCSC. We acknowledge the use of the following internet-based resources: the ESO Digitized Sky Survey; the TEPCat catalog; the SIMBAD data base operated at CDS, Strasbourg, France; and the arXiv scientific paper preprint service operated by Cornell University.
\end{acknowledgements}

\bibliographystyle{aa}

\end{document}